\documentclass[12pt]{iopart}
\usepackage{iopams} 
\usepackage{graphicx}
\usepackage{adjustbox} 
\usepackage{multirow} 
\usepackage{url}
\usepackage{setspace} 
\begin{document}

\title[The Last Meridian Circles of Pulkovo Observatory]{The Last Meridian Circles of Pulkovo Observatory}

\author{V. N. Yershov
}

\address{State University of Aerospace Instrumentation\\ 67 B.Morskaya, 190000, St. Petersburg;
previously at Pulkovo Observatory, 65(1) Pulkovskoye shosse, 196140 St. Petersburg, Russian Federation}
\ead{vyershov@guap.ru}
\vspace{10pt}
\begin{indented}
\item[]In blessed memory of the outstanding designer of telescopes Yu.S. Streletsky, who passed away untimely in January 2025.
\end{indented}

\begin{spacing}{0.8} 
\begin{abstract}
 Meridian circles played a fundamental role in astronomy since their invention in 1704. Then, at the end of the XX century, this function had been taken over by space astrometry, when the Hipparcos mission demonstrated the advantages of space astrometry by achieving the milliarcsecond level of accuracy. This historical sketch describes the development of the last meridian circles at Pulkovo Observatory (St. Petersburg) in the last quarter of the XX century.
\end{abstract}
\end{spacing}
%

\section{Introduction}

The meridian circle, invented by Ole R{\o}mer in 1704, became the main astronomical instrument for three centuries 
in many astronomical observatories around the world.

\begin{figure}[htb]
\hspace{2.4cm}
\includegraphics[scale=0.4]{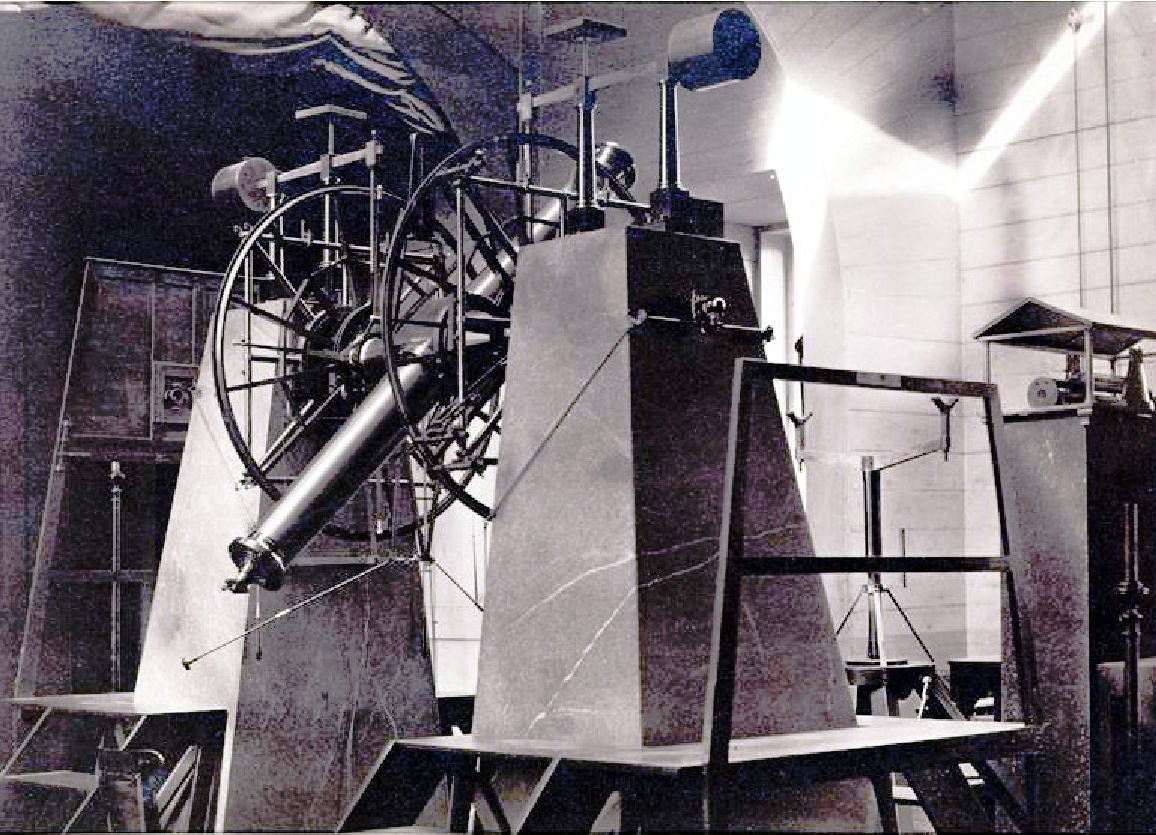}
\caption{Repsold meridian circle at Pulkovo Observatory in 1876 (photo from the Pulkovo Observatory archive)} 
\label{fig:repsold_mk}
\end{figure}

Therefore, in 1839, the founder and first director of Pulkovo Observatory near St. Petersburg, 
F.W. Struve, equipped the new observatory with a meridian circle ($D=150$\,mm, $F=2150$\,mm), which was ordered from
the Repsold company (Hamburg) and  which became one of the observatory's  main instruments, Figure\,\ref{fig:repsold_mk}.
 
\begin{figure}[htb]
%
\includegraphics[scale=0.4]{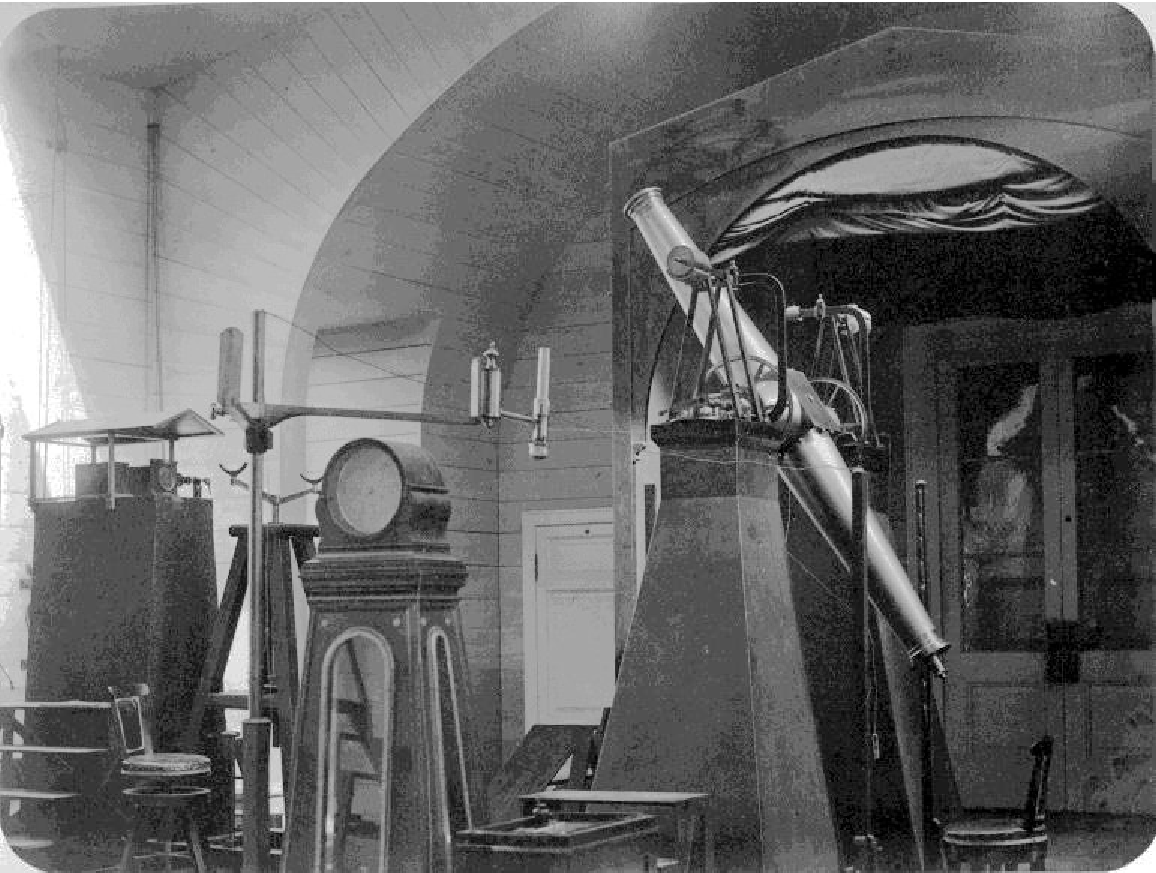}
%
\includegraphics[scale=0.4]{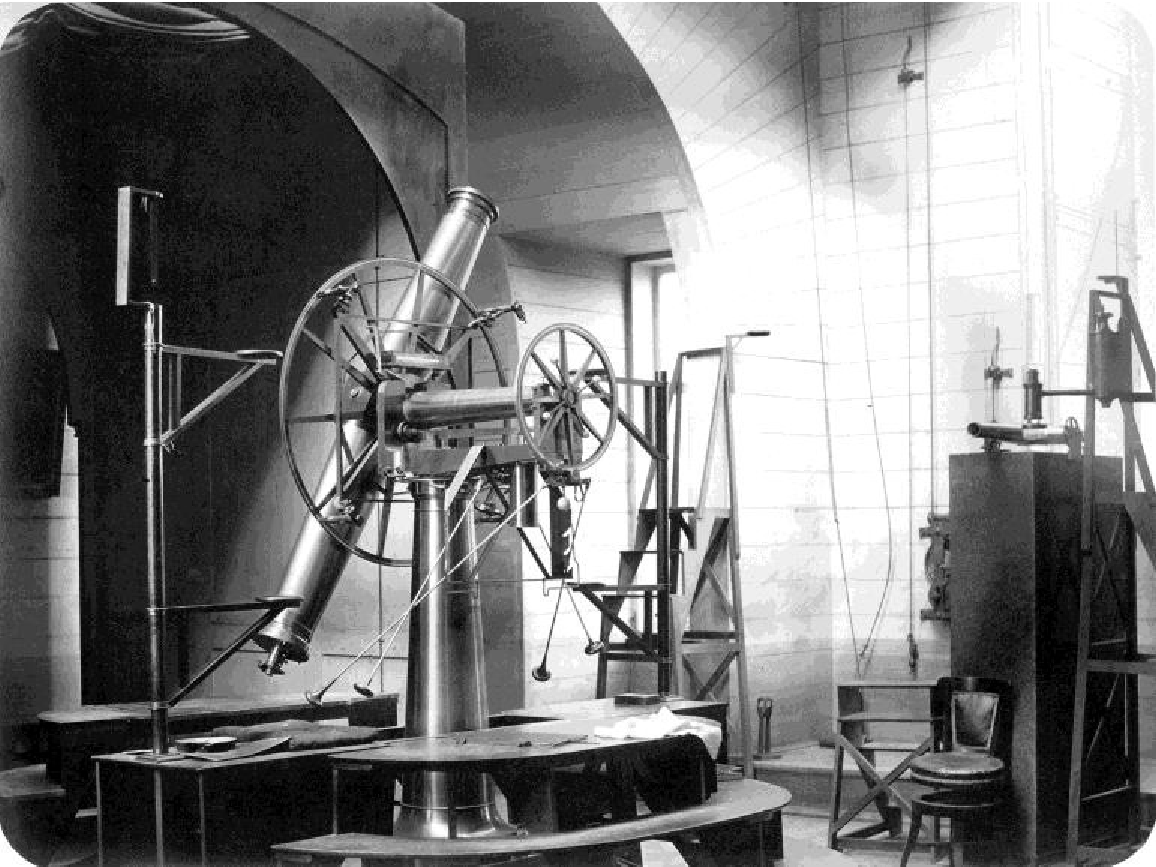}
\caption{The Ertel-Struve large transit instrument (left) and the Ertel-Struve 
vertical circle (right) in 1876 (photos from the Pulkovo Observatory archive).} 
\label{fig:ertel_transit}
\end{figure}

The meridian circle determines both right ascensions and declinations of stars passing through 
the meridian. However, Struve understood that higher accuracy could be achieved by measuring each 
coordinate separately. So, he ordered two other meridian instruments at a renowned German factory for precision 
mechanics lead by T.L. Ertel –- a large transit instrument ($D=150$~mm, $F=2590$~mm, Figure\,\ref{fig:ertel_transit}, left) 
and a vertical circle ($D=150$~mm, $F=1960$~mm, Figure\,\ref{fig:ertel_transit}, right). 
Struve participated in the design of these instruments as he wanted these instruments to be optimised for the 
absolute determinations of right ascensions and declinations.

 
\begin{figure}[htb]
\hspace{2.4cm}
\includegraphics[scale=0.35]{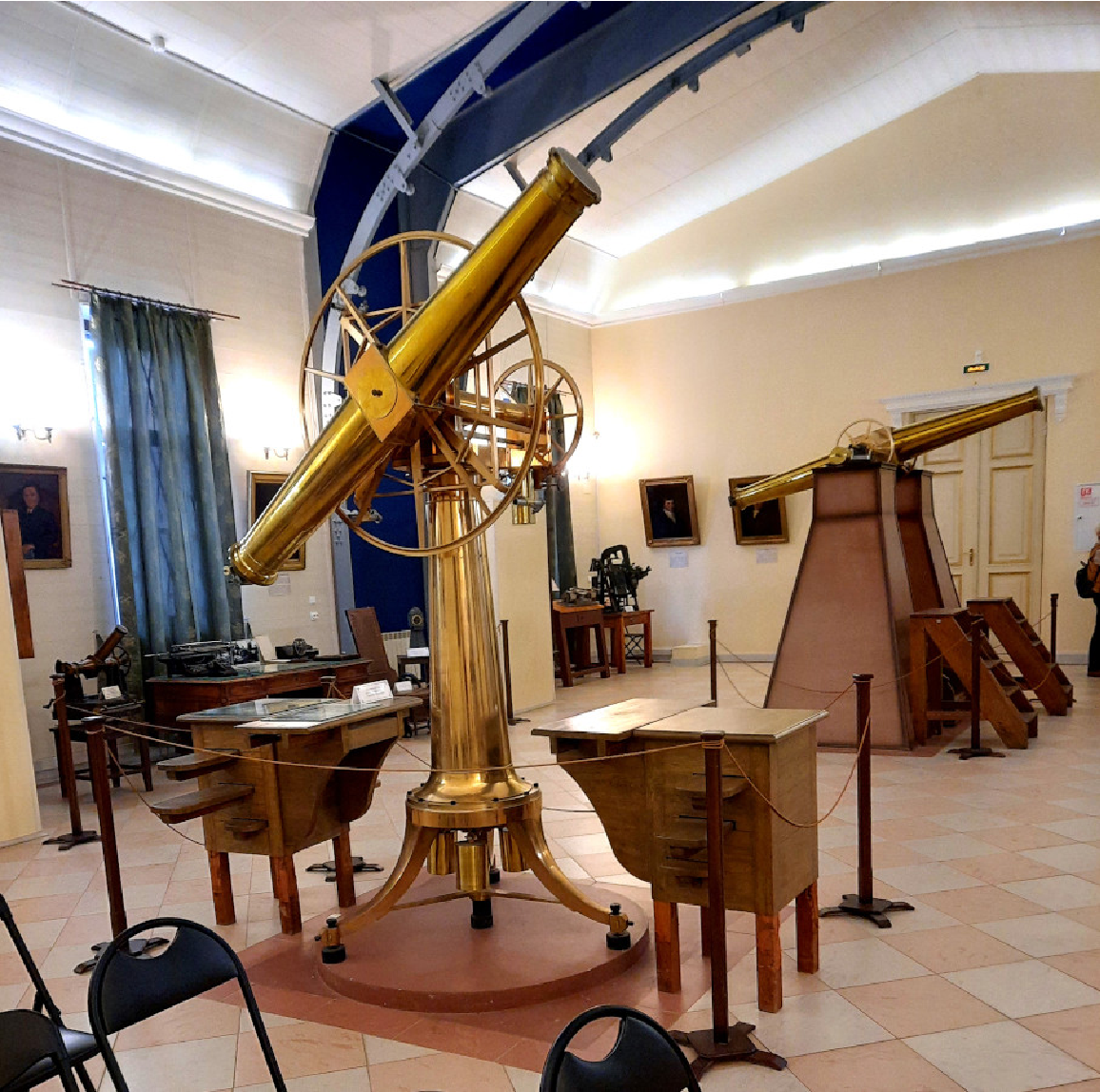}
\caption{Vertical circle and the large transit instrument by Ertel-Struve in the museum of
Pulkovo Observatory (2024, photo by the author)} 
\label{fig:ertel_instruments2024}
\end{figure}

\noindent
For a century, these instruments have produced measurements of the highest quality. 
But, since during the Second World War the front line passed through the territory of  Pulkovo Observatory, 
it was completely destroyed (only one small telescope dome on the southern 
slope of the Pulkovo Heights has survived).

The Repsold meridian circle was badly damaged during the war. In 1955, this instrument 
was transported to the Nikolaev branch of Pulkovo Observatory (now known as the Nikolaev 
Astronomical Observatory). After its restoration, the instrument
served for more than 40 years, producing a series of nine differential catalogs of 
star positions.

\section{Reconstruction of the classical meridian circle of Pulkovo Observatory}

\begin{figure}[htb]
\vspace{-0.3cm}
\hspace{2.4cm}
\includegraphics[scale=0.66]{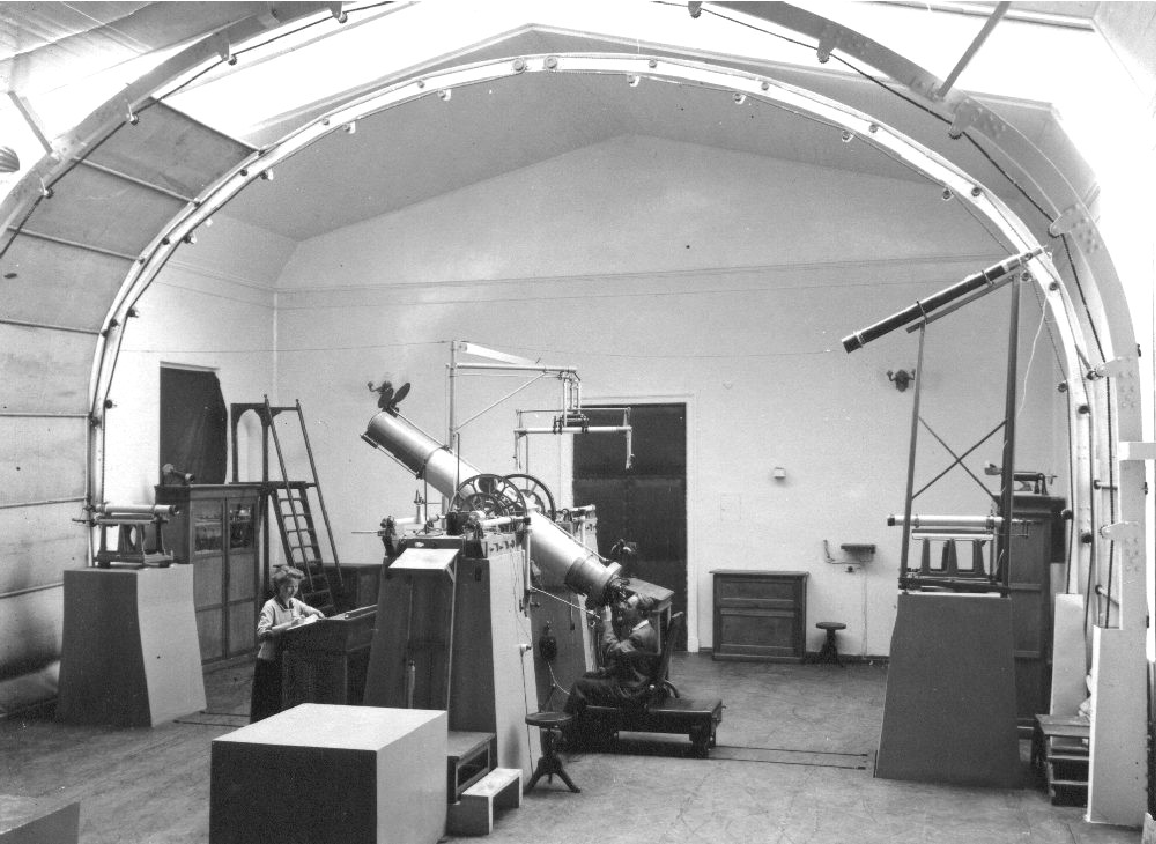}
\caption{The Toepfer meridian circle used at Pulkovo Observatory before 1978 -- installed in one of the 
meridian halls of Pulkovo Observatory (the observer M.S. Zverev and his assistant K.G. Gnevisheva, 
photo by A.F. Sukhonos, about 1974).} 
\label{fig:toepfer_mk}
\end{figure}

Pulkovo Observatory was reopened in 1954, at which time it received two large German telescopes 
as compensation: a 26-inch Zeiss refractor and a 190-mm Toepfer meridian circle (from Potsdam). 
The latter instrument ($D=190$\,mm, $F=2500$\,mm) was installed in one of the meridian halls 
of the observatory's main building where it was used until the late 1970s 
(Figure\,\ref{fig:toepfer_mk}) for classical visual observations of stars with the purpose
of producing differential catalogues of star positions.

In 1978, it was decided to reconstruct the instrument, providing it with photoelectric measuring devices for the 
focal plane and for  the graduated circle reading. The first stage of the reconstruction 
(the design of the focal-plane photoelectric micrometer and its controlling system) was finished 
in 1983 \cite{yershov84, yershov86}. Since the renewed instrument was supposed to use  
a photomultiplier as a detector, the optical system of the telescope was re-calculated for the blue wavelength 
band, and a new 200-mm diameter objective ($F=2000$\,mm) was manufactured. 

\begin{figure}[htb]
\hspace{2.4cm}
\includegraphics[scale=0.6]{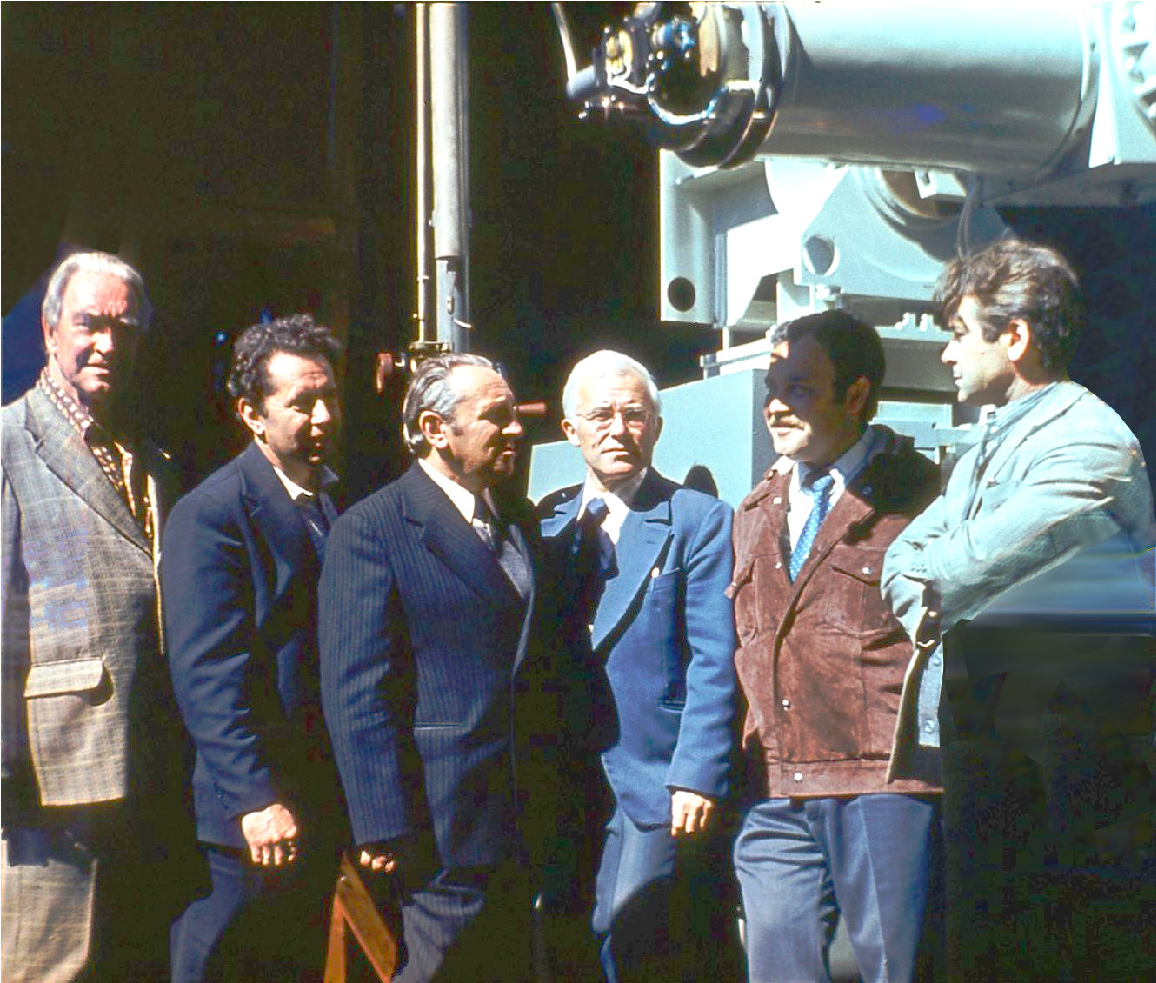}
\caption{Part of the  team reconstructing the Toepfer meridian circle in 1978-1980, from left to right:
M.S. Zverev, V.E. Pliss, K.N. Tavastsherna, Yu.S. Streletsky, E.G. Zhilisky, and  A.I. Schacht (photo by the author, 1982).} 
\label{fig:mk200_1982}
\end{figure}

\begin{figure}[htb]
\hspace{2.4cm}
\includegraphics[scale=0.4]{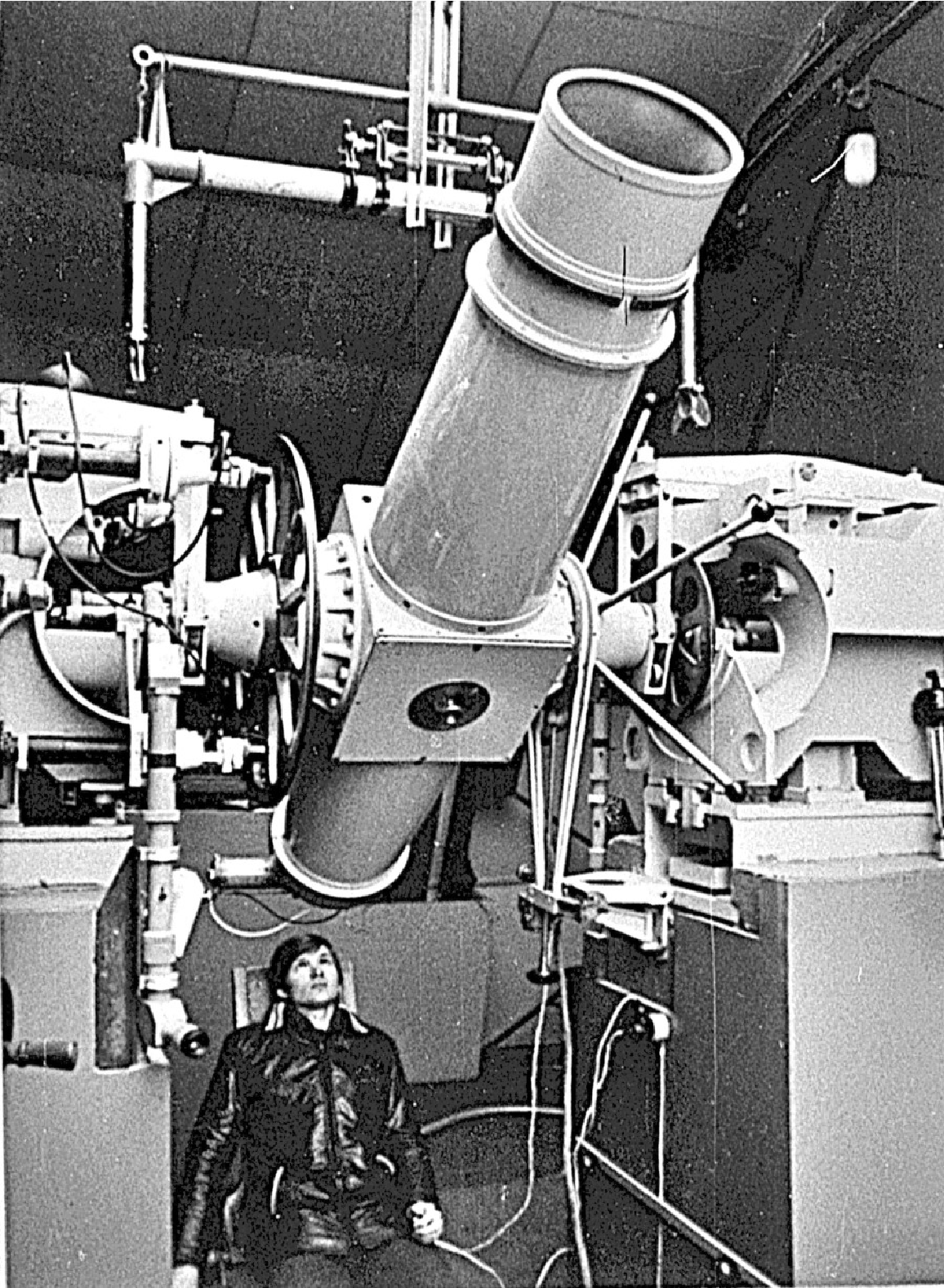}
\caption{The reconstructed Toepfer meridian circle (called MK-200) in 1984 (the observer V.N. Yershov, photo by A.F. Sukhonos).} 
\label{fig:mk200_1984}
\end{figure}

\begin{figure}[htb]
\hspace{2.4cm}
\includegraphics[scale=0.4]{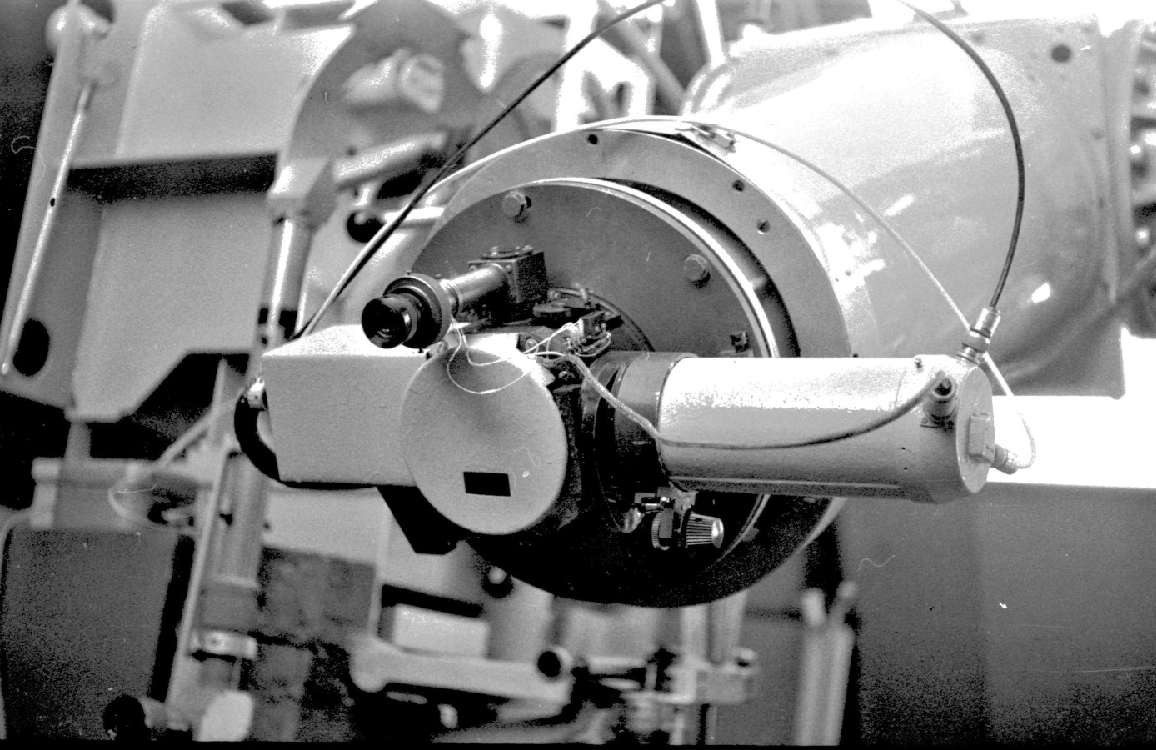}
\caption{Photoelectric micrometer of the MK-200 meridian circle (1986).} 
\label{fig:mk200_micrometer}
\end{figure}

The axis-support system was also completely re-designed (Figures\,\ref{fig:mk200_1982} and \ref{fig:mk200_1984}). 
In fact, it was a new instrument, equipped with a photoelectric focal-plane micrometer (Figure\,\ref{fig:mk200_micrometer}) 
and a new 200-mm objective calculated for the blue photoelectric wavelength band. So, it was renamed to be MK-200 
(highlighting its 200-mm diameter objective).  Its focal-plane micrometer had a V-shape slit pair, 
similar to that earlier proposed by E. H{\o}g \cite{hoeg71}, but these slits were not fixed: they were actively 
scanning a moving star image back and forth to obtain a number of transits of the star through the slits
(coordinates of the slits were measured by a digital angular encoder connected to the winding drive mechanism
of the carriage with the slits).
The relative velocity between the moving star image and
the slits was designed to be constant (independent on the star declination). This allowed observing 
slowly moving circumpolar stars and fixed meridian marks.   
  
The instrument did not have a driving mechanism to set it at the required zenith distance of the object under observation.
 Thus, it was a semi-automatic meridian circle, and the observer had to manually set the telescope at the zenith angle, 
 using the original Toepfer's setting graduated circle.
Then the observer could accurately adjust the telescope's zenith angle through a special ocular whose rotating 
prism was intercepting the field of view during the setting procedure. 
In this regime, the observer ensured a running star image to be within a narrow path marked in the ocular, 
which corresponded to the scanning path of the V-shaped slits. Then the prism was removed from the field of view, 
and the automatic registration begun.  

\begin{figure}[htb]
%
\includegraphics[scale=0.4]{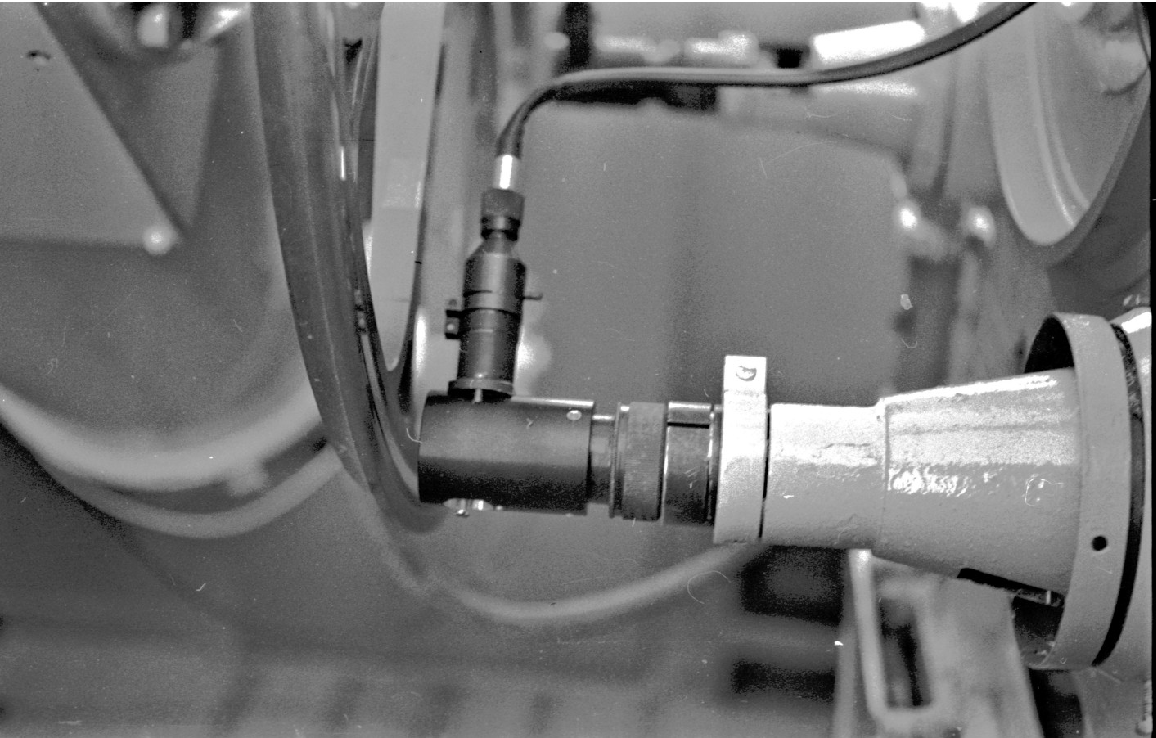}
%
\includegraphics[scale=0.4]{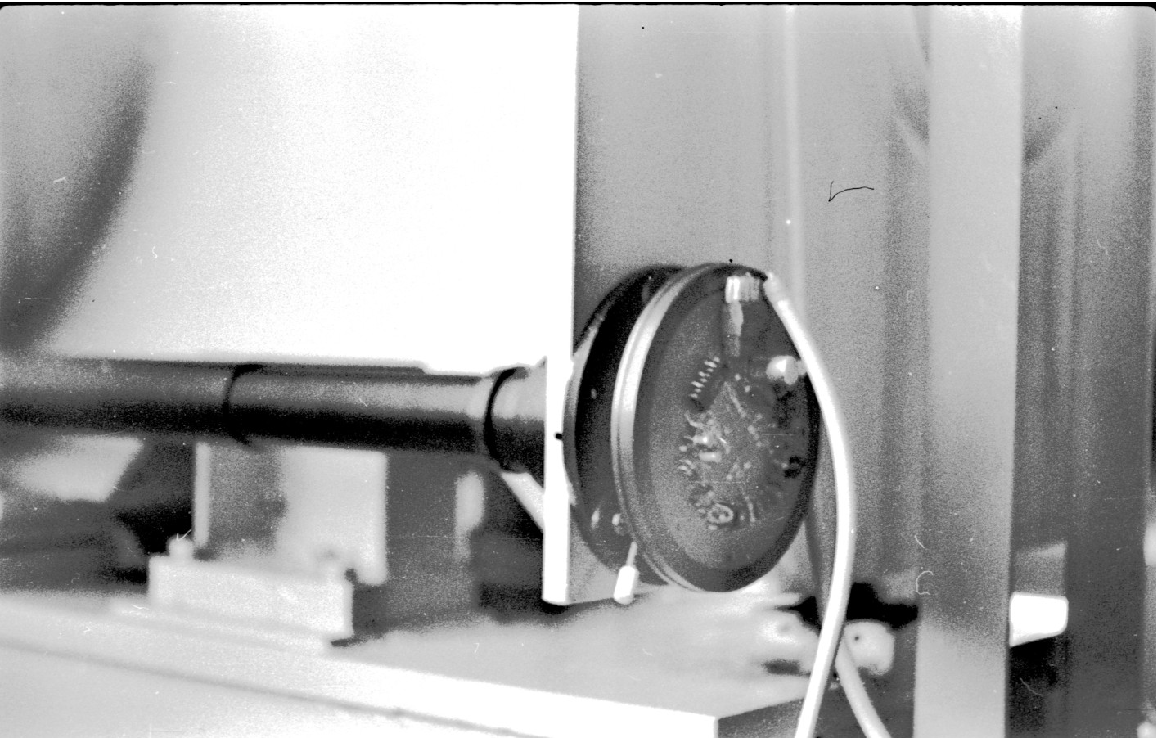}
\caption{The objective (left) and the CCD detector (right) of the MK-200 circle-reading system (photos by the author, 1986).} 
\label{fig:ccd_circle_reading}
\end{figure}

During the second stage of the reconstruction, a CCD-based photoelectric circle-reading detectors were manufactured 
(Figure\,\ref{fig:ccd_circle_reading})  which allowed the automatic reading of four microscopes. 
The circle reading procedure was simultaneous with the run of the scanning photoelectric focal-plane 
micrometer. 
The CCD devices for this system were manufactured by the Leningrad Research Institute of Television \cite{arutyunov86}.  

\begin{figure}[htb]
%
\includegraphics[scale=0.4]{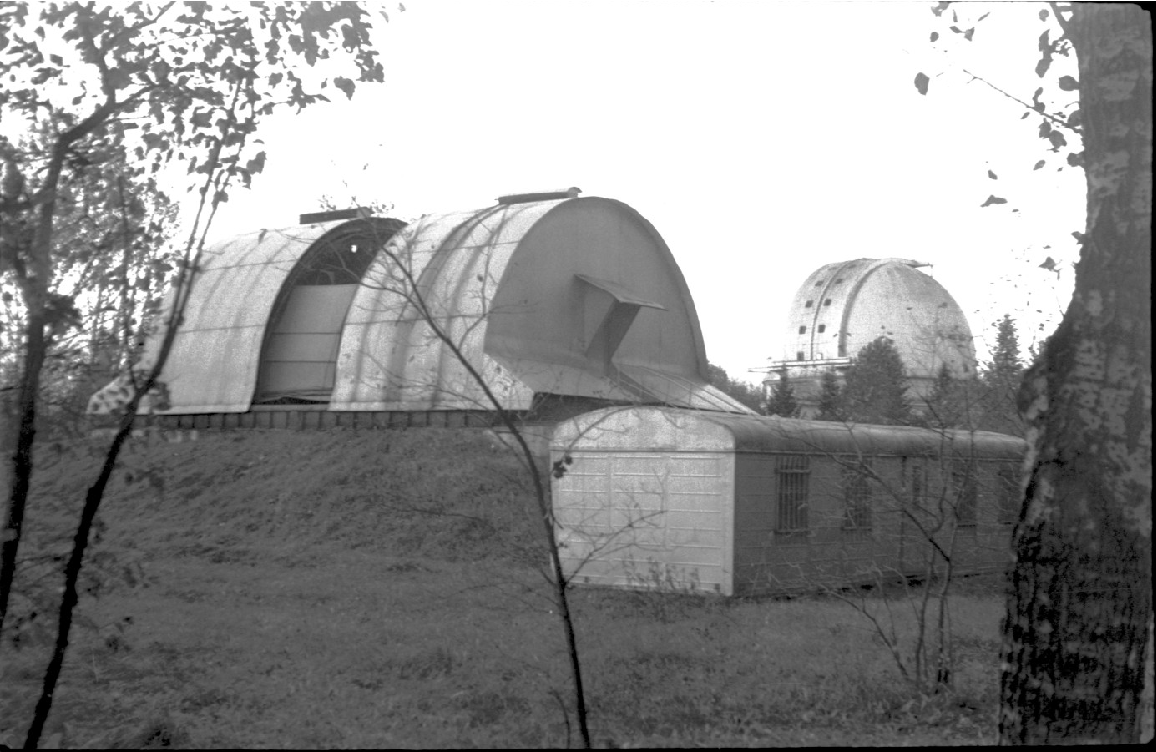}
%
\includegraphics[scale=0.4]{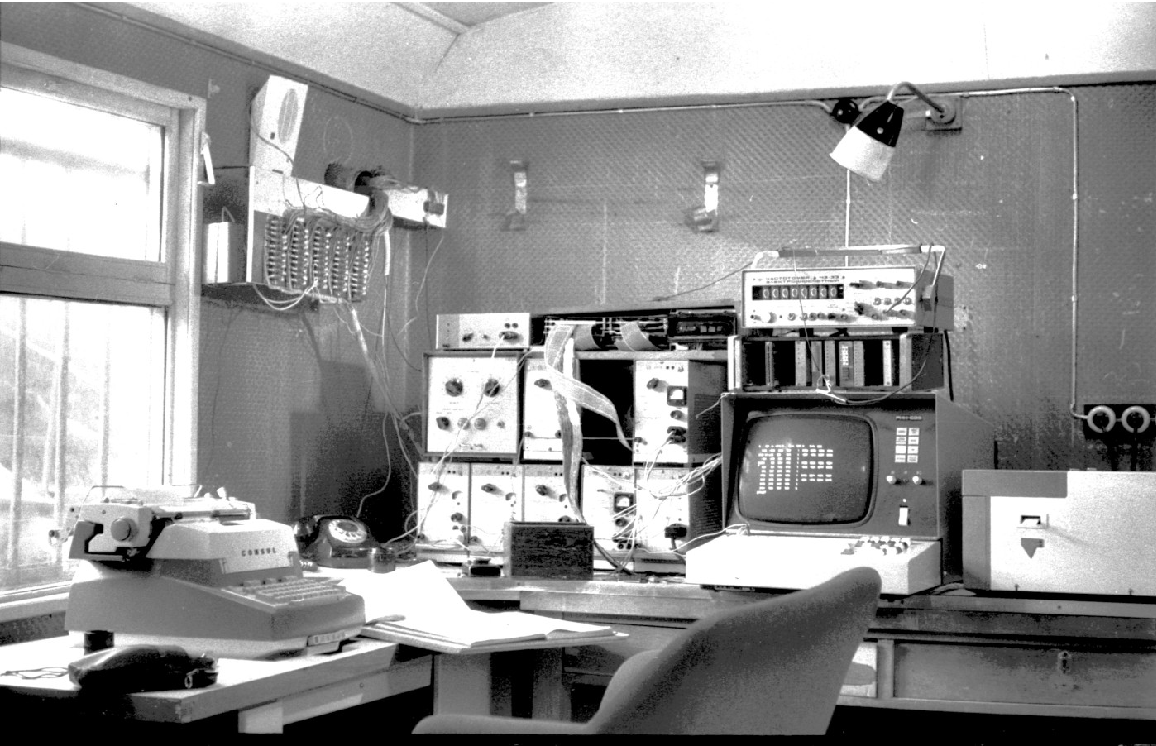}
\caption{The dome of the Pulkovo MK-200 meridian circle and the control room next to it (photos by the author, 1986).}
\label{fig:mk200_control_room}
\end{figure}

The  photoelectric focal-plane micrometer and the circle-reading CCDs were controlled
by logic chips and a 1980-style microcontroller (with punched tapes for loading software 
and for saving the digital outputs from the micrometer and CCDs). 
The control cabin of MK-200 (Figure\,\ref{fig:mk200_control_room})
The room-temperature cabin with the control devices was outside of the telescope dome 
(Figure\,\ref{fig:mk200_control_room}) to avoid systematic errors due to temperature gradients inside the dome. 

At the beginning of the 1990s, it was realised that meridian circles of classical type would soon become obsolete. 
Therefore, the astrometric instrumentation team focused its attention on a completely new meridian circle, which should be able 
to observe fainter stars in the infrared \cite{yershov95a,yershov95c}. 
The team members were K.N. Tavastsherna (the director of Pulkovo Observatory), M.S. Zverev and A.A. Nemiro
(the leading astrometrists), Yu.S. Streletsky (the leading telescope designer), V.E. Pliss (telescope designer),
G.O. Sokolov and Yu. G. Ostrensky (electronic engineers),  A.I. Schacht and Yu.N. Terkin (electronic and mechanics 
technicians), E.G. Zhilinky and the author of these notes (research astrometrists). Part of this team is shown in Figure\,\ref{fig:mk200_1982}.

Based on the experience obtained with the MK-200, the team was planning to create a giant near-infrared
meridian circle whose optical and mechanical elements were monitored by autocollimators. The eight optical axes
of the circle-reading auto-collimating microscopes were made in the form of spherical mirrors polished on the
laterals of a monolithic block of the primary mirror \cite{yershov95b}. The graduated circles 
read by the auto-collimating microscopes moving together with the primary mirror were fixed 
stationary on the side columns.
The central block of the instrument had also a spherical auto-collimating surface for monitoring the telescope's 
optical axis. All optical elements were made of a crystalline glass-ceramic with ultra-low coefficient of thermal expansion 
(called Astrositall) to avoid temperature deformations.  
The astrometric instrumentation group managed to manufacture the central block with its primary mirror and other 
optical elements, after which the collapse of the economy of the country did not allow to continue this work.

\section{Kislovodsk Mountain Station of Pulkovo Observatory}

\begin{figure}[htb]
%
\includegraphics[scale=0.4]{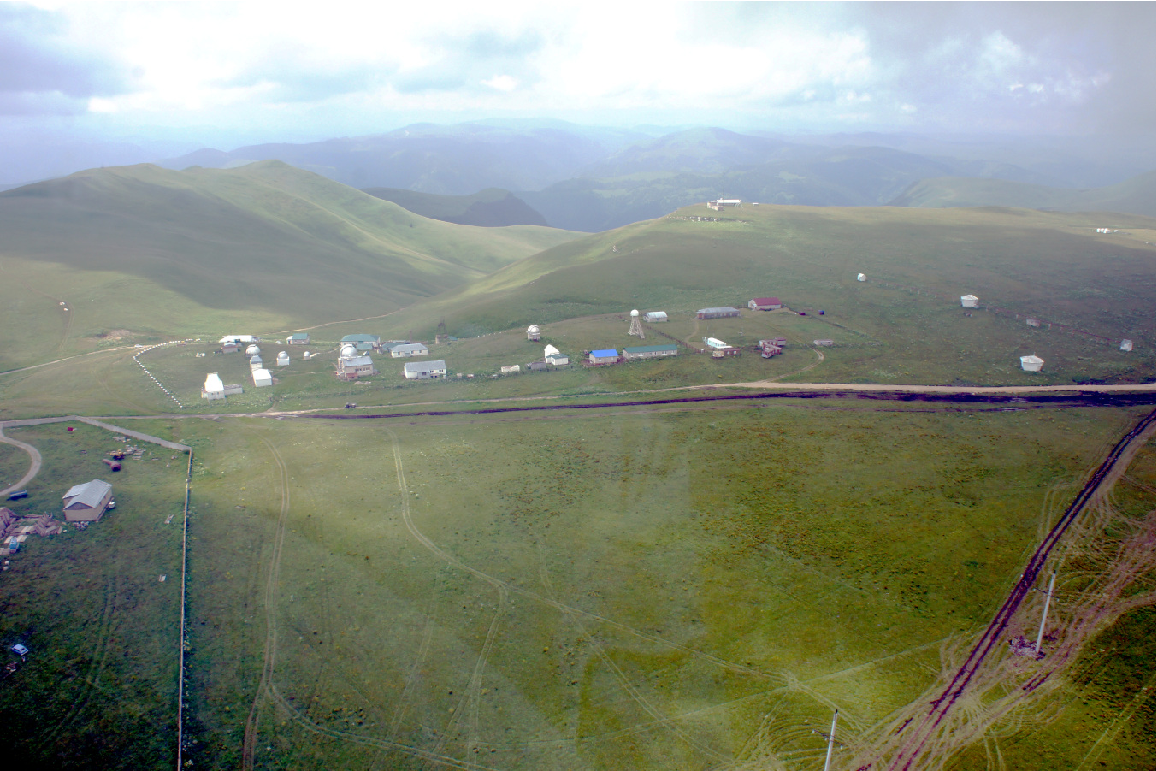}
%
\includegraphics[scale=0.4]{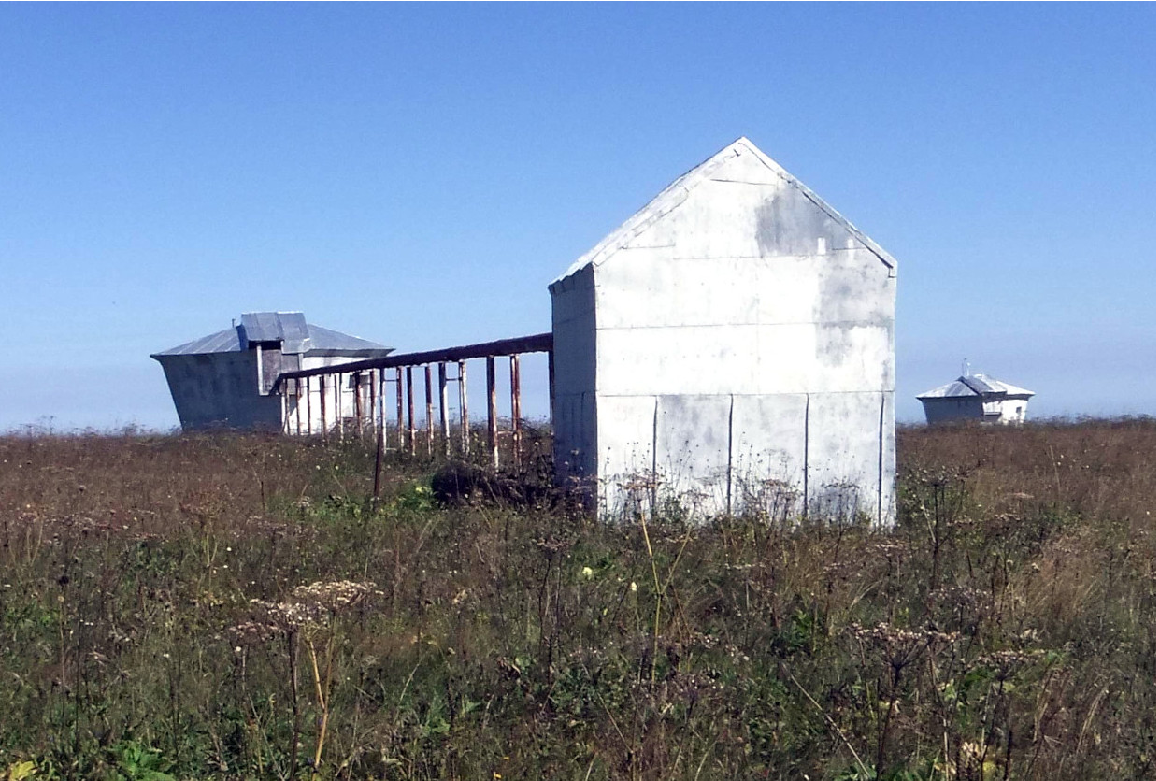}
\caption{Kislovodsk Mountain Station of Pulkovo Observatory (left) where two Ertel-Struve classical instruments, the vertical circle and the large transit instrument were installed during 1984 to 1989 in the rhomboidal-shape domes (right) designed by Yu.S. Streletsky. One of the domes is equipped with two meridian marks for monitoring the transit instrument’s azimuth 
(photos by the author, 2012).} 
\label{fig:kislovodsk_domes2022}
\end{figure}

There were no meridian circles at the Kislovodsk Mountain Station of Pulkovo Observatory
(Figure\,\ref{fig:kislovodsk_domes2022}, left). 
This observation site was built at the altitude of 2100\,m by a leading Pulkovo solar physicist 
M.N. Gnevyshev in 1948 (before the reopening of Pulkovo Observatory). The goal of this site was 
to regular monitor the solar corona by using coronagraphic telescopes. The astronomical climate of this site
was (and is) good, and in the 1980s it was decided to move there some of the ground-based astrometric 
observations.
   
\begin{figure}[htb]
%
\includegraphics[scale=0.4]{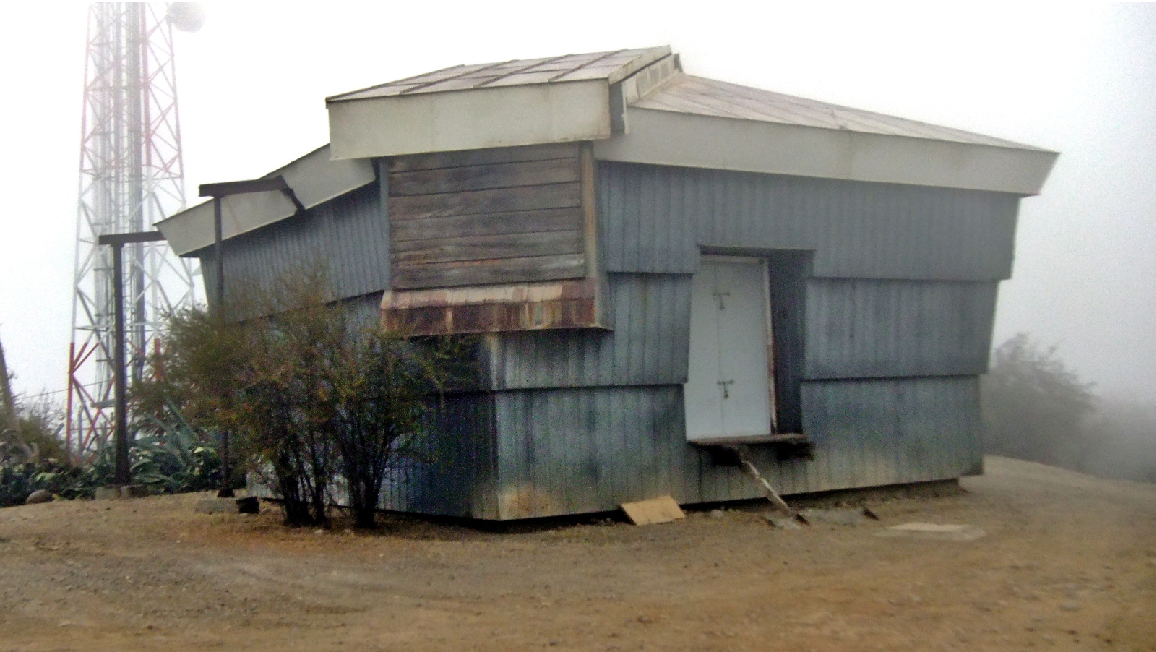}
%
\includegraphics[scale=0.4]{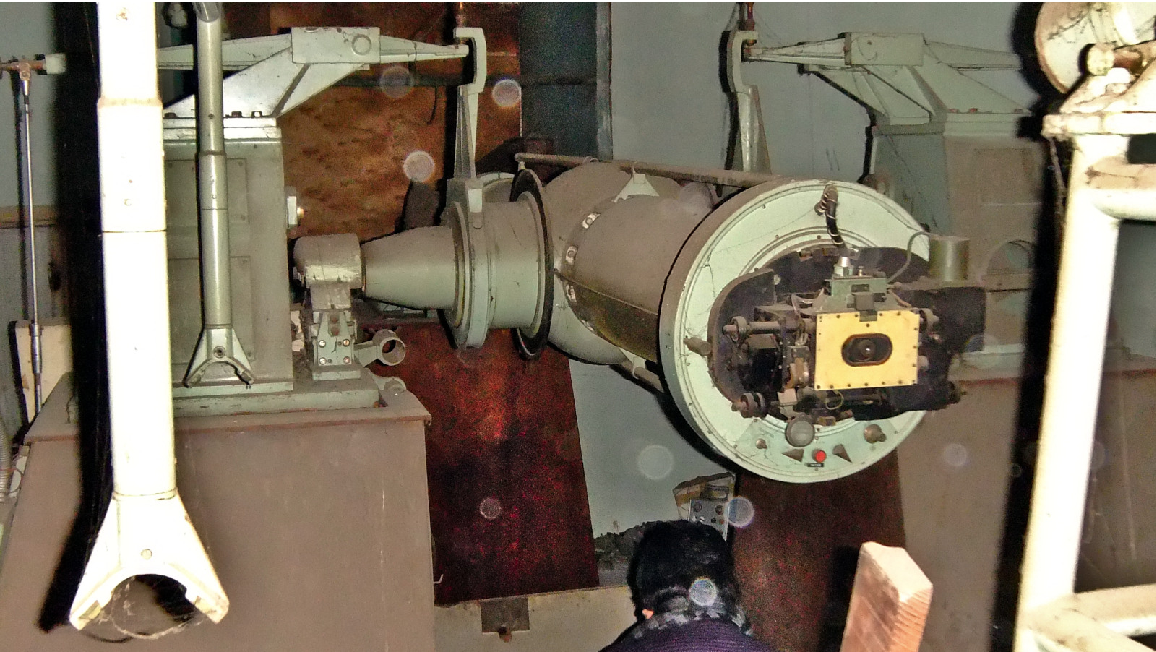}
\caption{The rhomboidal-shape dome designed by Yu.S. Streletsky for the Pulkovo large transit instrument
at the Cerro Calan observatory (Chile) in the 1960s and the large transit instrument inside (photos
by the author, 2013).} 
\label{fig:cerro_calan_dome2013}
\end{figure}

The testing of this site was based on the use of two old classical instruments acquired by Struve in XIX century 
from Ertel's company. These instruments were installed at the Kislovodsk Mountain Station 
in 1984 in two rhomboidal-shaped domes (Figure\,\ref{fig:kislovodsk_domes2022}, right) 
constructed by an outstanding Pulkovo telescope designer Y.S. Streletsky. 
He used his original concept of automatic temperature maintenance in pavilions of astrometric telescopes, 
which he applied in the 1960s, while constructing a pavilion for a large transit instrument 
installed at the observation site of the Cerro Calan National Astronomical  Observatory, Chile
(Figure\,\ref{fig:cerro_calan_dome2013}). In that work, Pulkovo maintained 
Struve's idea of observing right ascensions and declinations separately. Thus, two instruments
were used at Cerro Calan: the large transit instrument for right ascensions and Zverev's 
photographic vertical circle for declinations.  
Similarly, at the Kislovodsk Mountain Station, the two Ertel-Struve instruments were installed 
for separately determining right ascensions and declinations of Solar System bodies.  
The observations lasted from 1984 to 1989 and resulted in 566 declinations of the Sun, 230 of Mercury, 413 of Venus, 
and 207 of Mars. Similar numbers of right ascensions were obtained by the Ertel-Struve 
large transit instrument. 
Observations were visual, and their reductions were differential in the FK5 system.  
Only day-time observations of reference stars were used \cite{devyatkin90,devyatkin09}. 
After finishing these classical-style observations, the two instruments were returned 
back to the Pulkovo Observatory museum (Figure\,\ref{fig:ertel_instruments2024}).

\section{Sukharev Horizontal Meridian Circle }

The Pulkovo horizontal meridian circle \cite {kanaev10,pinigin00} has undergone changes 
in its formation for at least three periods. It was conceived by its author, L.A. Sukharev, 
who studied a concept of this instrument in the 1952-53 by using its scaled models.

\begin{figure}[htb]
\hspace{2.0cm}
%
%
\includegraphics[scale=0.35]{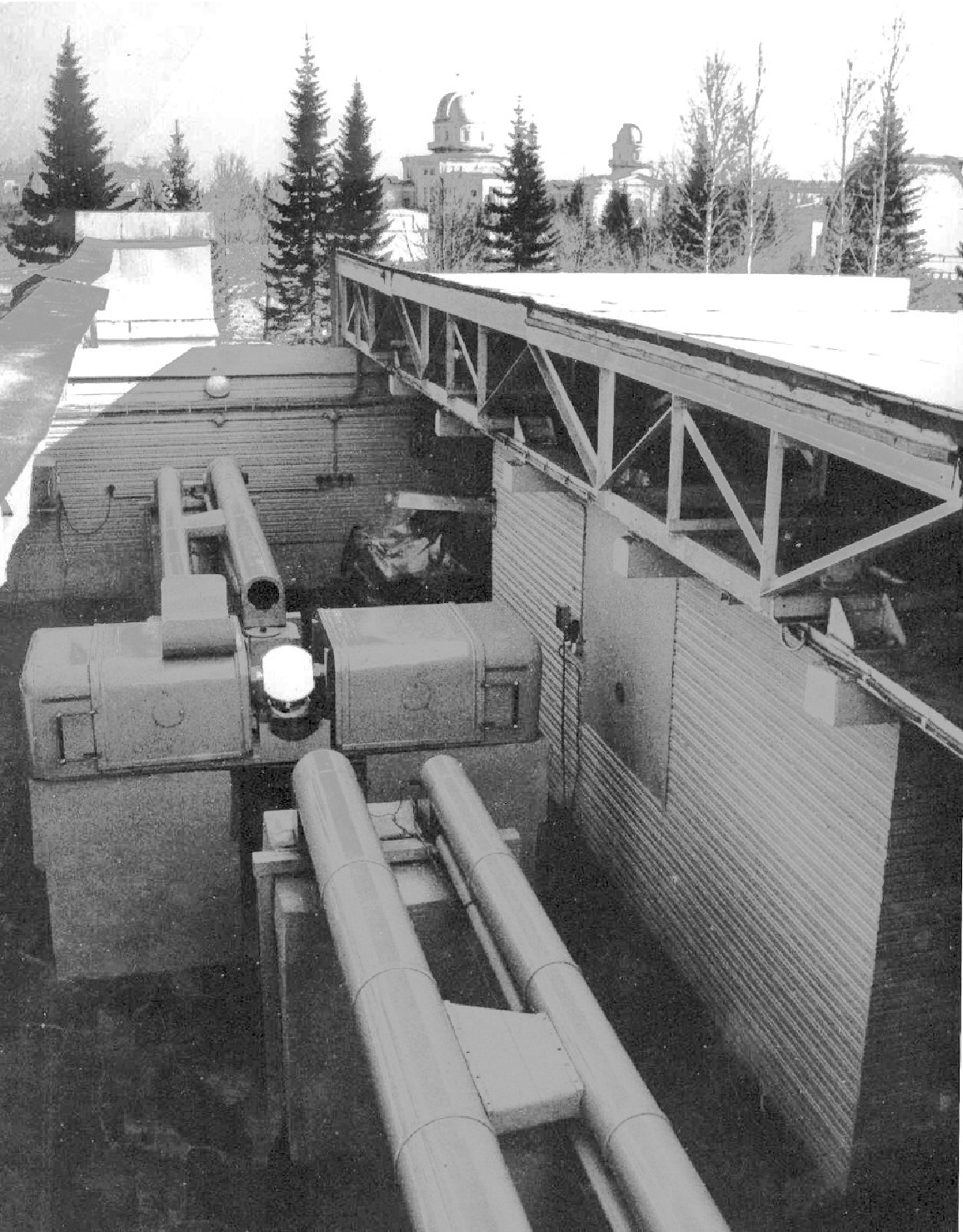}
\includegraphics[scale=0.35]{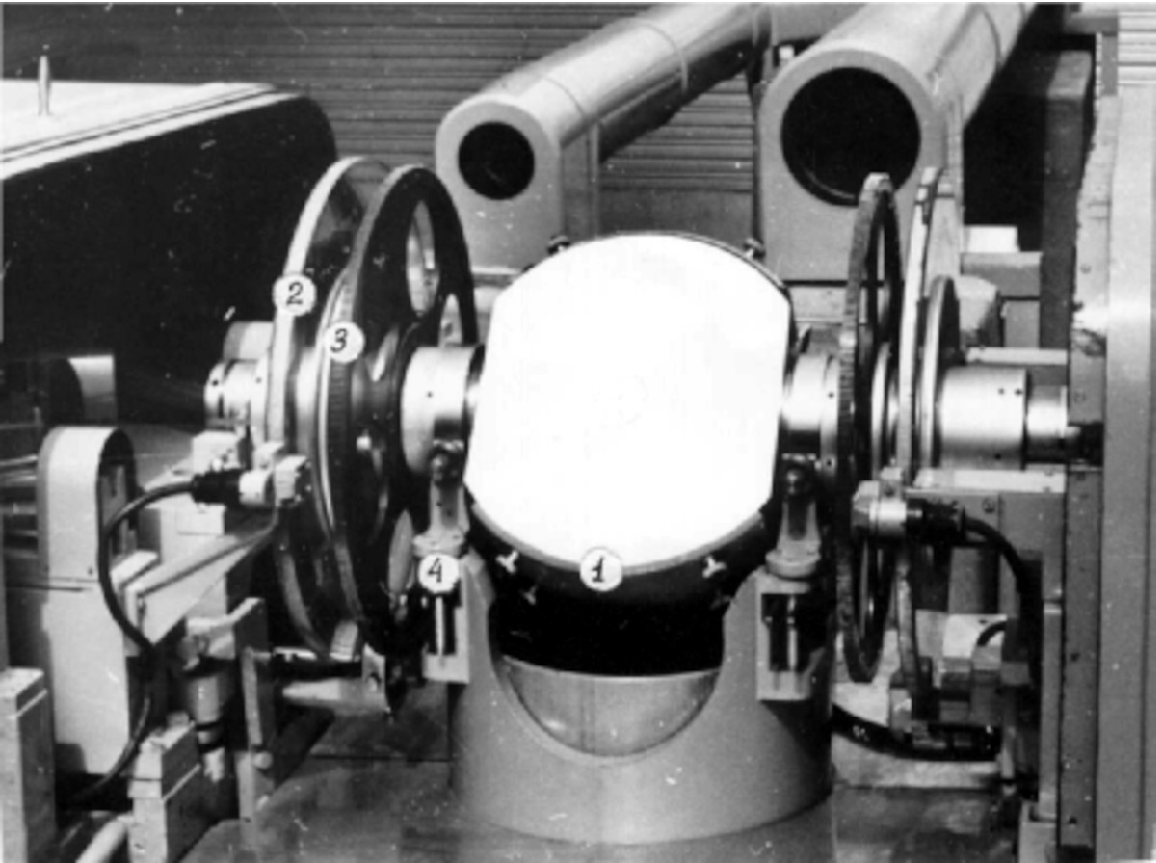}
\caption{Sukharev horizontal meridian circle of Pulkovo Observatory.
General view  (left) and its central part (right):  1 -- a flat mirror; 2 -- a graduated circle; 
3 –- a gear, 4 –- an unloading system (photos by A.F. Sukhonos, 1978).} 
\label{fig:horizontal_merkrug1}
\end{figure}

In 1960, a horizontal meridian circle made at the Kiev ``Arsenal'' factory and named by L.A. Sukharev \cite{pinigin00} 
was installed at Pulkovo (Figure\,\ref{fig:horizontal_merkrug1}). After the refinement of the photoelectric ocular micrometer 
and other units, in the spring of 1967, the study of this instrument was started by G.I. Pinigin who used the FK4 catalogue. 
In the subsequent years, with the active participation of Kazan astronomers R.I. Gumerov, V.B. Kapkov and others, high-precision recording devices were developed for the Sukharev Horizontal Meridian Circle, and an electronic control system was also created.

In the 1980s, after the development and study of the electronic control system, regular observations of bright and faint stars at both coordinates were carried out at the Sukharev Horizontal Meridian Circle in the automatic regime, in order to obtain a catalogue of positions of the weak part of the FK5, as well as a differential catalogue of positions of reference stars located in sites with radio sources. 

In the 1990s, a cooperation between Pulkovo Observatory, Nikolaev Observatory and 
the Engelhardt Observatory of the Kazan University (I.I. Kanaev, G.I. Pinigin, A.V. Shulga, 
O.E. Shornikov, A.V. Sergeev) led to the creation of a series of four automatic horizontal 
meridian circles based on the Sukharev's design \cite{kanaev10}. They were active until the early 2000s, 
determining the differential right ascensions and declinations of objects up to 17th magnitude 
in the B, V, and R bands of the Johnson photometric system.

\section{Conclusions}

With these observations, the epoch of meridian circles had finished at Pulkovo Observatory. 
Differential astrometric observations continue using several instruments, both at the scientific 
site in Pulkovo (the telescope ZA-320M) and at the Kislovodsk Mountain Astronomical Station
(the telescope MTM-500M). The main goal of these observations is to monitor  
asteroids and other cosmic bodies appearing in near-Earth space \cite{devyatkin16}.

The fundamental coordinate system which was previously maintained by meridian circles 
is currently based on distant extragalactic radio sources observed by using the very-long-baseline 
radio interferometric technique. 
The propagation of this fundamental system to other coordinate systems is carried out by 
space astrometry missions, such as Gaia. This work is (and will be) continued 
by other space and ground-based means. Instead of meridian circles, an optimal ground-based instrument 
might be a wide-field telescope with one- or two-meter primary mirror  and with an astrometric-quality
field of view of a few dozen square degrees, like that of the one-meter AZT-16 Maksutov  
telescope installed at the El Roble site of the National Astronomical Observatory, Chile 
\cite{mikhelson76,deviatkin15, alonso22}. 
Still, the best quality observations with the telescopes of this kind will be obtained in the meridian.

\end{document}